%% file: indriskpaper.tex
\documentclass[12pt]{article}
\usepackage{a4wide,latexsym,xspace,oupbib} \usepackage{hyperref}
\input{bknot} 
\renewcommand{\secref}[1]{\mbox{\S\ref{sec:#1}}}

\title{On Individual Risk}

\author{Philip Dawid\\University of Cambridge, UK}

\begin{document}
\maketitle

\begin{abstract}
  We survey a variety of possible explications of the term
  ``Individual Risk.''  These in turn are based on a variety of
  interpretations of ``Probability,'' including Classical,
  Enumerative, Frequency, Formal, Metaphysical, Personal,
  Propensity, Chance and Logical conceptions of Probability, which we
  review and compare.  We distinguish between ``groupist'' and
  ``individualist'' understandings of Probability, and explore both
  ``group to individual'' (G2i) and ``individual to group'' (i2G)
  approaches to characterising Individual Risk.  Although in the end
  that concept remains subtle and elusive, some pragmatic suggestions
  for progress are made.
\end{abstract}

\tableofcontents
\clearpage

\section{Introduction}
\label{sec:intro}

``Probability'' and ``Risk'' are subtle and ambiguous concepts,
subject to a wide range of understandings and interpretations.  Major
differences in the interpretation of Probability underlie the
fundamental Frequentist/Bayesian schism in modern Statistical Science.
However, these terms, or their synonyms, are also in widespread
general use, where lack of appreciation of these subtleties can lead
to ambiguity, confusion, and outright nonsense.  At the very least,
such usages deserve careful attention to determine whether and when
they are meaningful, and if so in how many different ways.  The focus
of this article will be on the concept of ``Individual Risk,'' which I
shall subject to just such deep analysis.

To set the scene, \secref{ex} presents some examples displaying a
variety of disparate usages of the concept of Individual Risk.  These,
in turn, are predicated on a variety of different understandings of
the concept of Probability.  I survey these in \secref{interp},
returning to discuss the examples in their light in \secref{revisit}.
Section~\ref{sec:expert} describes the concept of ``expert
assignment,'' which is is common to a number of understandings of
Risk.  In \secref{g2i} and \secref{indrisk} I consider various aspects
of ``group to individual'' (G2i) inference---the attempt to make sense
of Individual Risk when Probability is understood as a group
phenomenon---and conclude that this can not be done in an entirely
satisfactory way.  So \secref{i2g} reverses this process and considers
``individual to group'' (i2G) inference, where we take Individual Risk
as a fundamental concept, and explore what that implies about the
behaviour of group frequencies.  The i2G approach appears to lead to
essentially unique individual risk values; but these can not be
considered absolute, but are relative to a specified information base,
and moreover are typically not computable.  In \secref{disc} I take
stock of the overall picture.  I conclude that the concept of
``Individual Risk'' remains highly problematic at a deep philosophical
level, but that does not preclude our making some pragmatically
valuable use of that concept---so long as we are aware of the various
pitfalls that may lie in our path.

\section{Examples}
\label{sec:ex}
We start with a m\'enagerie of examples of ``individual risk'' in
public discourse.

\begin{ex}
  \label{ex:weather}
  A weather forecaster appears on television every night and issues a
  statement of the form ``The probability of precipitation tomorrow is
  $30\%$'' (where the quoted probability will of course vary from day
  to day).  Different forecasters issue different probabilities.
\end{ex}

\begin{ex}
  \label{ex:arai}
  There has been much recent interest concerning the use of Actual
  Risk Assessment Instruments (ARAIs): statistical procedures for
  assessing ``the risk'' of an individual becoming violent \cite{BSL}.
  Thus a typical output of the Classification of Violence Risk (COVR)
  software program, that can be used to inform diagnostic testimony in
  civil commitment cases, might be: ``The likelihood that XXX will
  commit a violent act toward another person in the next several
  months is estimated to be between 20 and 32 percent, with a best
  estimate of 26 percent.''
\end{ex}

\begin{ex}
  \label{ex:aharoni}
  \textcite{aharoni:2013} tested a group of released adult offenders
  on a go/no-go task using fMRI, and examined the relation between
  task-related activity in the anterior cingulate cortex (ACC) and
  subsequent rearrest (over four years), allowing for a variety of
  other risk factors.  They found a significant relationship between
  ACC activation on the go/no-go task and subsequent rearrest; whereas
  subjects with high ACC activity had a 31\% chance of rearrest,
  subjects with low ACC activity had a 52\% chance.  They conclude:
  ``These results suggest a potential neurocognitive biomarker for
  persistent antisocial behavior.''

  A newly released offender has low ACC activity: how should we judge
  his chance of rearrest?
\end{ex}

\begin{ex}
  \label{ex:jolie}
  Writing in the
  \href{http://www.nytimes.com/2013/05/14/opinion/my-medical-choice.html}{New
    York Times (May 14 2013)} about her decision to have a preventive
  double mastectomy, the actress Angelina Jolie said: ``I carry a
  faulty gene, BRCA1, which sharply increases my risk of developing
  breast cancer and ovarian cancer.  My doctors estimated that I had
  an 87 percent risk of breast cancer and a 50 percent risk of ovarian
  cancer, although the risk is different in the case of each woman.''
\end{ex}

\begin{ex}
  \label{ex:climate}
  The
  \href{http://www.ipcc.ch/news_and_events/docs/ar5/ar5_wg1_headlines.pdf}{Fifth
    Assessment Report of the Intergovernmental Panel on Climate
    Change}, issued in September 2013, contains the statements ``It is
  {\em extremely likely\/} that human influence has been the dominant
  cause of the observed warming since the mid-20th century,'' ``It is
  {\em virtually certain\/} that the upper ocean ($0$--$700$ m) warmed
  from 1971 to 2010, and it {\em likely\/} warmed between the 1870s
  and 1971,'' and ``It is {\em very likely\/} that the Arctic sea ice
  cover will continue to shrink and thin and that Northern Hemisphere
  spring snow cover will decrease during the 21st century'' (their
  italics).
  \href{http://www.climatechange2013.org/images/uploads/WGIAR5-SPM_Approved27Sep2013.pdf}{It
    is explained that} {\em virtually certain\/} is equivalent to a
  probability of at least $99\%$, {\em extremely likely\/}, at least
  $95\%$, {\em very likely\/}, at least $90\%$, and {\em likely\/}, at
  least $66\%$.
\end{ex}

\begin{ex}
  \label{ex:obama}
  On 4 May 2011, three days after he announced that American troops
  had killed Osama bin Laden in Pakistan, US President Barack Obama
  said in an
  \href{http://www.cbsnews.com/8301-504803_162-20060530-10391709.html}{interview}
  with ``60 Minutes'' correspondent Steve Kroft:
  \begin{quote}
    At the end of the day, this was still a 55/45 situation.  I mean,
    we could not say definitively that bin Laden was there.
  \end{quote}
\end{ex}

\begin{ex}
  \label{ex:civil}
  In a civil court case, the judgment might be expressed as: This is
  typically interpreted as ``with probability exceeding $50\%$.''
\end{ex}

In all the above examples we can ask: How were the quoted
probabilities interpreted?  How might they be interpreted?  And how
might the quality of such probability forecasts be measured?

% \begin{itemize}
% \item DJS book

% \item : ``a stochastic model for individual risk.''  Also related to
%   my work with Carlo!

% \item And what is status of Rubin's PO apporach?  Outcomes without
%   risks?

% \end{itemize}

\section{Interpretations of Probability}
\label{sec:interp}

We have already remarked that the concept of ``Probability'' is a
hotly contested philosophical issue.\footnote{Interestingly, this is
  not the case for its mathematical properties, where the ``Kolmogorov
  axioms'' are largely accepted by all the main factions.}  Even the
many who have no patience for such philosophising are usually in
thrall to some implicit philosophical conception, which shapes their
approach and understanding, and their often fruitless arguments with
others who (whether or not so recognised) have a different
understanding.

One axis along which the different theories and conceptions of
Probability can be laid out---and which is particularly germane to our
present purpose---is whether they regard Probability as fundamentally
an attribute of groups, or of individuals.  I will refer to these as,
respectively, ``groupist'' and ``individualist'' theories.

Below, in a necessarily abbreviated and admittedly idiosyncratic
account, I outline some of the different conceptions of Probability,
and hope to bring out their relationships, similarities and
differences.  For a fuller discussion see \eg\
\textcite{hajek:sep,galavotti:2014}.

\subsection{Classical Probability}
\label{sec:classprob}
If you studied any Probability at school, it will have focused on the
behaviour of unbiased coins, well-shuffled packs of cards, perfectly
balanced roulette wheels, \etc, \etc\@ In short, an excellent training
for a life misspent in the Casino.  This is the ambit of {\em
  Classical Probability\/}.

The underlying conception is that we have a number of {\em elementary
  outcomes\/} of an experiment, exactly one of which will actually be
realised when the experiment is performed.  For example, there are
$N=53,644,737,765,488,792,839,237,440,000$ ways in which the cards at
Bridge can be distributed among 4 players, and just one of these ways
will materialise when the cards are dealt.  Any event of interest, for
example ``North holds 3 aces,'' can be represented by the set of all
the elementary outcomes for which it is the case; and the number $n$
of these, divided by the total number $N$ of all elementary outcomes,
is taken as the measure of the probability of the event in question.
The mathematics of Classical Probability is thus really a branch of
Combinatorial Analysis, the far from trivial mathematical theory of
counting.

Since the focus is on the specific outcome of (say) a particular deal
of cards or roll of a die, this classical conception is individualist.
But questions as to the interpretation of the ``probabilities''
computed rarely raise their heads.  If they do, it would typically be
assumed (or regarded as a precondition for the applicability of the
theory) that each of the elementary outcomes is just as likely as any
other to become the realised outcome.  However, in the absence of any
independent understanding of the meaning of ``likely,'' such an
attempt at interpretation courts logical circularity.  Furthermore,
paradoxes arise when there is more than one natural way to describe
what the elementary events should be.  If we toss two coins, we could
either form 3 elementary events: ``0 heads,'' ``1 head,'' ``2 heads'';
or, taking order into account, 4 elementary events: ``tail tail,''
``tail head,'' ``head tail,'' ``head head.''  There is nothing within
the theory to say we should prefer one choice over the other.  Also,
it is problematic to extend the classical conception to cope with an
infinite number of events: for example, to describe a ``random
positive integer,'' or a line intersecting a given circle ``at
random.''

The principal modern application of Classical Probability is to
situations---for example, clinical trials---where randomisation is
required to ensure fair allocation of treatments to individuals.  This
can be effected by tossing a ``fair coin'' (or by simulating such
tosses on a computer).

\subsection{Enumerative Probability}
\label{sec:enumprob}

What I~\footnote{Though probably no one else.  \textcite{hajek:sep}
  uses the term ``finite frequentism.''} here term {\em Enumerative
  Probability\/} can also be regarded as an exercise in counting.
Only now, instead of counting elementary outcomes of an experiment, we
consider a finite collection of individuals (of any nature), on which
we can measure one or more pre-existing attributes.

Thus consider a set of individuals, $I_1, I_2, \ldots, I_N$, and an
attribute $E$ that a generic individual $I$ may or may not possess:
\eg\ ``smoker.''  A specific individual $I_k$ can be classified
according to whether or not its ``instance,'' $E_k$, of that attribute
is present---\eg, whether or not $I_k$ is a smoker.  If we knew this
for every individual, we could compute the {\em relative frequency\/}
with which $E$ occurs in the set, which is just the number of
individuals having attribute $E$, divided by the total number $N$ of
individuals.  This relative frequency is the ``enumerative
probability'' of $E$ in the specified set.  Clearly this is a
``groupist'' conception of probability.

Of course, the value of such an enumerative probability will depend on
the set of individuals considered, as well as what attributes they
possess.  We will often care about the constitution of some specific
large population (say, the population of the United Kingdom), but have
data only for a small sample.  Understanding the relationship between
the known frequencies in the sample and the unknown frequencies in the
whole population is the focus of the theory and methodology of sample
surveys.  This typically requires the use of randomisation to select
the sample---so relying, in part, on the Classical conception.

Although based on similar constructions in terms of counting and
combinatorial analysis, Classical and Enumerative Probability are
quite different in their scope and application.  Thus suppose we ask:
What is ``the probability'' that a new-born child in the United
Kingdom will be a boy?  Using the Classical interpretation, with just
two outcomes, boy or girl, the answer would be $1/2 = 0.5$.  But in
the UK population about $53\%$ of live births are male, so the
associated Enumerative Probability is $0.53$.

\subsection{Frequency Probability}
\label{sec:freqprob}

``Frequency probability'' can be thought of as enumerative probability
stretched to its limits: instead of a finite set of individuals, we
consider an infinite set.

There are two immediate problems with this:
\begin{enumerate}
\item \label{it:prob1} Infinite populations do not exist in the real
  world, so any such set is an idealisation.
\item \label{it:prob2} Notwithstanding that the great statistician Sir
  Ronald Fisher manipulated ``proportions in an infinite population''
  with great abandon and to generally good effect, this is not a
  well-defined mathematical concept.
\end{enumerate}

These problems are to some extent resolved in the usual scenario to
which this concept of probability is attached: that of {\em repeated
  trials under identical conditions\/}.  The archetypical example is
that of tosses of a possibly biased coin, supposed to be repeated
indefinitely.  The ``individuals'' are now the individual tosses, and
the generic outcome is (say) heads (H), having instances of the form
``toss $i$ results in heads.''  It is important that the
``individuals'' are arranged in a definite sequence, for example time
order.  Then for any finite integer $N$ we can restrict attention to
the sequence of all tosses from toss 1 up to toss $N$, and form the
associated enumerative probability, $f_N$ say, of heads: the relative
frequency of heads in this finite set.  We next consider how $f_N$
behaves as we increase the total number $N$ without limit.  If $f_N$
approaches closer and closer to some mathematical limit $p$---the
``limiting relative frequency'' of heads---then that limiting value
may be termed the ``frequency probability'' of heads in the sequence.

Of course we can never observe infinitely many tosses, so even the
existence of the limit must remain an assumption: one that is,
however, given some empirical support by the observed behaviour of
real-world frequencies in such repeated trial situations.  But even
when we can happily believe that the limit $p$ exists, we will never
have the infinite amount of data that would be needed to determine its
value precisely.  Much of the enterprise of {\em statistical
  inference\/} addresses the subtle relationship between actual
frequencies, observed in finite sequence of trials, and the ideal
``frequency probabilities'' that inhabit infinity.

Richard~\textcite{vonmises} attempted to build a sophisticated
mathematical theory of probability built upon the above frequency
conception as its primitive element, but that is now largely of
historical interest.

\subsection{Formal Probability}
\label{sec:statprob}
A prime task for an applied statistician is to build a ``statistical
model'' of a phenomenon of interest.  That phenomenon could be fairly
simple, such as recording the outcomes of a sequence of 10 tosses of a
coin; or much more sophisticated, such as a description of the earth's
temperature as it varies over space and time.  Any such model will
have symbols representing particular outcomes and quantities (\eg, the
result of the 10th toss of the coin, or the recorded temperature at
the Greenwich Royal Observatory at noon GMT on 24 September 2020), and
will model these (jointly with all the other outcomes and quantities
under consideration) as ``random variables,'' having a joint
probability distribution.  Typically the probabilities figuring in
this distribution are treated as unknown, and statistical
data-analysis is conducted to learn something about them.  There does
not seem to be a generally accepted terminology for these
probabilities figuring in a statistical model: I shall term them
``Formal Probabilities.''

Since the focus of statistical attention is on learning these formal
probabilities, I find it remarkable how little discussion is to be
found as to their meaning and interpretation.  In particular, is
Formal Probability an individualist or a groupist conception?  The
elementary building blocks of the statistical model are assignments of
probabilities to specific events, so {\em prima facie\/} it looks like
an individualist conception.  One possible individualist
interpretation of a formal probability value is as a propensity---see
\secref{propensity} below---although this raises problems of its own.

In any case, many users of statistical models would not be happy to
accept an individualist interpretation of their formal probabilities,
and prefer a groupist account of them.  Thus consider the simple case
of coin-tossing.  The usual statistical model for this assigns some
common (though typically unknown) formal probability $p$ to each event
$E_k$: ``the $k$th toss $I_k$ results in heads,'' for every $k$; and
further models all these tosses, as $k$ varies, as independent.  An
application of Bernoulli's Law of Large Numbers shows that this
implies that, with probability 1, the limiting relative frequency of
heads in the sequence will exist and have value $p$.  And many
statisticians would claim that---notwithstanding that the formal
probabilities are attached to individual tosses---this groupist
Frequency Probability interpretation is the {\em only\/} meaningful
content of the (superfically individualist) formal ``Bernoulli
model.''

However, such a ploy is far from straightforward for more complex
models, where there may be no natural way of embedding an individual
event in a sequence of similar events.  For example, if we construct a
spatio-temporal statistical model of the weather, the model might well
contain a parameter that measures the correlation between the
temperatures in Greenwich and New York at noon GMT on 24 September
2020.  But just how is such a correlation, between two ``one-off''
quantities, to be interpreted?  Does it have any real-world
counterpart?

One possible way of taking the above ``Law of Large Numbers'' approach
for the Bernoulli model and extending it to more general models
\cite{apd:synthesis} is by identifying those events that are assigned
probability 1 by the assumed model, and asserting that the {\em
  only\/} valid interpretation of the model lies in its claim that
these events will occur.  But this may be seen as throwing too much
away.

Because there is little discussion and no real shared understanding of
the interpretation of Formal Probability, pointless disagreements can
spring up as to the appropriateness of a statistical model of some
phenomenon.  An example of practical importance for forensic DNA
profiling arises in population genetics, where there has been
disagreement as to whether the genes of distinct individuals within
the same subpopulation should be modelled as independent, as claimed
by \textcite{Roeder}, or correlated, as claimed by
\textcite{for/smith}.  However, without a shared understanding of what
(if anything) the correlation parameter in the formal model relates to
in the real world, this is a pointless argument.  It was pointed out
by \textcite{apd:isba} that (like blind men's disparate understandings
of the same elephant) both positions can be incorporated within the
same hierarchical statistical model, where the formal correlation can
come or go, according to what it is taken as conditioned on.  So while
the two approaches superficially appear at odds, at a deeper level
they are in agreement.

% A statistical model for ``individual risks'' that calls for deeper
% interpretation of this term is that of \textcite{jr+sg:89}, which
% elaborates the ``potential outcome'' approach to causal inference
% \cite{dbr:as} by modelling individual hazard rates, rather than
% individual outcomes, as random.  See \secref{comm} below for one
% attempt at such interpretation.

\subsubsection{Metaphysical Probability}
\label{sec:meta}
There are those who, insistent on having some sort of frequency-based
foundation for probability, would attempt to interpret a ``one-off''
probability as an average over repetitions of the whole underlying
phenomenon.  For example, the correlation between the temperatures in
Greenwich and New York at noon GMT on 24 September 2020 would be taken
to refer to an average over---necessarily hypothetical---independent
repetitions of the whole past and future development of weather on
Earth.  This conception appears close to the ``many worlds''
interpretation of quantum probabilities currently popular with some
physicists.  I confess I find it closer to science fiction than to
real science---``metaphysical,'' not ``physical.''  At any rate, since
we can never observe beyond the single universe we in fact inhabit, we
can not make any practical use of such hypothetical repetitions.

That said, this way of thinking can be useful, purely as an analogy,
in helping people internalise a probability value (whatever its
provenance or philosophical back-story).  Thus one of the options at\\
\href{http://understandinguncertainty.org/files/animations/RiskDisplay1/RiskDisplay.html}{\small
  \tt
  http://understandinguncertainty.org/files/animations/RiskDisplay1/RiskDisplay.html}\\
for visualising a risk value is a graphic containing a number of
symbols for your ``possible futures,'' marked as red to indicate that
the event occurs in that future, green that it does not.  The relevant
probability is then represented by the proportion of these symbols
that are red.  It is not necessary to take the ``possible futures''
story seriously in order to find this enumerative representation of a
probability value helpful.  The same purely psychological conceit
explains the misleading name given to the so-called ``frequentist''
approach to statistical inference.  This involves computing the
probabilistic properties of a suggested statistical procedure.  For
example, a test of a null hypothesis at significance level $5\%$ is
constructed to have the property that probability of deciding to
reject the hypothesis, under the assumption that it is in fact
correct, is at most $0.05$.  Fisher's explication of such a test was
that, when the it results in the decision to reject, either the null
hypothesis was indeed false, or else an event of small probability has
happened---and since we can largely discount the latter alternative,
we have evidence for the former.  Although for this purpose it would
be perfectly satisfactory to interpret the probability value $0.05$ as
an ``individual chance,'' relating solely to the current specific
application of the procedure, it is almost universally expressed
metaphysically, by a phrase such as ``over many hypothetical
repetitions of the same procedure, when the null hypothesis is true,
it will rejected at most $5\%$ of the time.''

\subsection{Personal Probability}
\label{sec:persprob} {\em Personal\/} (often, though less
appropriately, termed {\em Subjective\/}) {\em Probability\/} is very
much an individualist approach.  Indeed, not only does it associate a
probability value, say $p$, with an individual event, say $E$, it also
associates it with the individual, say ``You,'' who is making the
assignments, as well as (explicitly or implicitly) with the
information, say $H$, available to You when You make the assignment.
Thus if You are a weather forecaster, You might assess Your
probability of rain tomorrow, given Your knowledge of historical
weather to date, at $40\%$.  Another forecaster would probably make a
different personal assessment, as would You if You had different
information (perhaps additional output from a meteorological computer
system) or were predicting for a different day.

What is the meaning of Your $40\%$?  The standard view is that it
simply describes the odds at which You would be willing to bet on
``rain tomorrow.''  If Classical Probability is modelled on the
casino, then Personal Probability is modelled on the race-track.

If You really really had to bet, You would have to come up with a
specific numerical value for Your personal probability.  Thus to a
Personalist there is no such thing as an ``unknown probability,'' and
it appears that these should not feature in any probability model You
might build.  There is however an interesting and instructive
relationship between such Personalist models and the statistical
models, as described in \secref{statprob}, that do feature unknown
formal probability values---for more on this see \secref{persprob2}
below.

\subsection{Propensity and chance}
\label{sec:propensity}
The interpretation of Probability as a ``Propensity'' was championed
by \textcite{popper:propen}, and is still much discussed by
philosophers---though hardly at all by statisticians.  The overall
idea is that a particular proposed coin-toss (under specified
circumstances) has a certain (typically unknown) ``propensity'' to
yields heads, if it were to be conducted---just as a particular lump
of arsenic has a propensity to cause death if it were to be ingested.
This is clearly an individualist (though non-Personalist) conception
of Probability, but there is little guidance available as to how a
propensity probability is to be understood, or how its value might be
assessed, except by reference to some ``groupist'' frequency
interpretation.  Indeed, some versions of the propensity account
\cite{gillies:bjps2000} take them as referring directly to the
behaviour of repetitive sequences, rather than to that of individual
events.

Another individualist term that is very close in spirit to
``propensity'' is ``(objective) chance.''  The ``Principal Principle''
of \textcite{lewis:80}, while not defining chance, relates it to
personal probability by requiring that, if You learn (don't ask how!)
that the chance of an event $A$ is (say) $0.6$, and nothing else, then
Your personal probability of $A$ should be updated to be $0.6$.  And
further conditions may be required: for example, that this assessment
would be unaffected if You learned of any other ``admissible'' event,
where ``admissible'' might mean ``prior to $A$,'' or ``independent of
$A$ in their joint chance distribution.''  Such conditions are
difficult to make precise and convincing \cite{pettigrew:2012}.  In
any case, the Principal Principle gives no guidance on how to compute
or estimate an individual chance.

\subsection{Logical Probability}
\label{sec:logprob}
Yet another non-Personalist individualist conception of Probability is
Logical Probability, associated with \textcite{jeffreys:39,carnap} and
others.  This is similar to the propensity/chance account in
considering that there is an ``objective'' probability value
associated with a given outcome, but differs from those in emphasising
the relativity of that value to the information upon which it is
premissed.  Once again there would seem to be no routine and
unambiguous way of computing such logical probability values.

\section{Examples Revisited}
\label{sec:revisit}
We here revisit the examples of \secref{ex} in the light of some of
the above discussion.  My own comments are interpolated in [\dots].

\setcounter{expl}{0}

\begin{ex}
  \label{ex:weather2}
  \textcite{gigerenzer:riskanal} randomly surveyed pedestrians in five
  metropolises located in countries that have had different degrees of
  exposure to probabilistic forecasts: Amsterdam, Athens, Berlin,
  Milan, and New York.  Participants were told to imagine that the
  weather forecast, based on today's weather constellation, predicts
  ``There is a $30\%$ chance of rain tomorrow,'' and to explain what
  they understood by that.

  \begin{quotation}
    Several people in New York and Berlin thought that the rain
    probability statement means ``3 out of 10 meteorologists believe
    it will rain'' [a form of enumerative probability?].  A woman in
    Berlin said, ``Thirty percent means that if you look up to the sky
    and see 100 clouds, then 30 of them are black'' [a different form
    of enumerative probability].  Participants in Amsterdam seemed the
    most inclined to interpret the probability in terms of the amount
    of rain. ``It's not about time, it indicates the amount of rain
    that will fall,'' explained a young woman in Amsterdam.  Some
    people seemed to intuitively grasp the essence of the ``days''
    interpretation, albeit in imaginative ways. For instance, a young
    woman in Athens in hippie attire responded, ``If we had 100 lives,
    it would rain in 30 of these tomorrow'' [a metaphysical
    interpretation?].  One of the few participants who pointed out the
    conflict between various interpretations observed, ``A probability
    is only about whether or not there is rain, but does not say
    anything about the time and region,'' while another said, ``It's
    only the probability that it rains at all, but not about how
    much'' [two individualist views]. Many participants acknowledged
    that, despite a feeling of knowing, they were incapable of
    explaining what a probability of rain means.

    [According to the authors of the paper, the standard
    meteorological interpretation is: when the weather conditions are
    like today, in 3 out of 10 cases there will be (at least a trace
    of) rain the next day.]
  \end{quotation}
\end{ex}

\begin{ex}
  \label{ex:arai2}
  There has been heated recent debate centred on the construction and
  validity of confidence intervals for the ``individual risks'' output
  by an ARAI.  This has been initiated and promoted by
  \textcite{HartMichieCooke,CookeMichie,HartCooke}, whose analysis has
  had a strong influence (see \eg\ \textcite{BSL}), but has been
  widely criticised for serious technical statistical errors and
  confusions
  \cite{HarrisRiceQuinsey,HansonHoward,imreydawid,mossman:2014}.
  However, to date that debate has largely sidestepped the crucial
  question of the meaning of such an individual risk.
    
  \textcite{HartMichieCooke} make the following argument in an attempt
  to distinguish between group and individual risks:
  \begin{quotation}
    To illustrate our use of Wilson's method for determining group and
    individual margins of error, let us take an example.  Suppose that
    Dealer, from an ordinary deck of cards, deals one to Player. If
    the card is a diamond, Player loses; but if the card is one of the
    other three suits, Player wins.  After each deal, Dealer replaces
    the card and shuffles the deck.  If Dealer and Player play 10 000
    times, Player should be expected to win $75\%$ of the time.
    Because the sample is so large, the margin of error for this group
    estimate is very small, with a $95\%$ CI of $74$--$76\%$ according
    to Wilson's method.  Put simply, Player can be $95\%$ certain that
    he will win between $74$ and $76\%$ of the time. However, as the
    number of plays decreases, the margin of error gets larger. If
    Dealer and Player play 1000 times, Player still should expect to
    win $75\%$ of the time, but the $95\%$ CI increases to
    $72$--$78\%$; if they play only 100 times, the $95\%$ CI increases
    to $66$--$82\%$. Finally, suppose we want to estimate the
    individual margin of error. For a single deal, the estimated
    probability of a win is still $75\%$ but the $95\%$ CI is
    $12$--$99\%$.  The simplest interpretation of this result is that
    Player cannot be highly confident that he will win---or lose---on
    a given deal
  \end{quotation}

  [On 10 000, 1000 or 100 deals, the actual success rate will vary
  randomly about its target value of $75\%$, and the ``confidence
  intervals'' (CIs) described for these cases are intended to give
  some idea of the possible extent of that variation.  But on a single
  deal the actual success rate can only be $0$ (which will be the case
  with probability $25\%$) or $100\%$ (with probability $75\%$).  This
  purely binary variation is not usefully described by any
  ``confidence interval,'' let alone the above one of $12$--$99\%$,
  based on a misconceived application of Wilson's method.  See
  \textcite{imreydawid} for further deconstruction of the spurious
  philosophical and mathematical arguments presented by these
  authors.]
\end{ex}

\begin{ex}
  \label{ex:aharoni2}
  \textcite{neurolaw} point out that the results given by
  \textcite{aharoni:2013} provide inflated estimates of the predictive
  accuracy of the model when generalizing to new individuals: their
  reanalysis using a cross-validation approach found that addition of
  brain activation to the predictive model improves predictive
  accuracy by less than $3\%$ compared to a baseline model without
  activation.  The alternative bootstrap-based reanalysis in
  \textcite{aharoni:2014} likewise finds the original estimates to
  have been over-inflated, though their correction is smaller.  The
  message is that it can be extremely difficult to obtain reliable
  estimates of reoffending rates, and any causal attribution, \eg\ to
  ACC activity, would be even more precarious.

  All these objections aside, suppose we had reliable estimates of the
  statistical parameters in a well-fitting statistical model, from
  which we could compute a ``reoffending rate'' for this new
  individual.  Is it appropriate to regard such a formal probability
  value, based on a statistical analysis of group data, as the
  appropriate expression of his individual risk?
\end{ex}

\begin{ex}
  \label{ex:jolie2}
  Angelina Jolie's statement sounds very much like an
  ``individualist'' interpretation of the figure ``87 percent risk of
  breast cancer,'' especially in the light of ``the risk is different
  in the case of each woman.''  However it would appear that this
  figure is taken from the website of
  \href{https://www.myriad.com/products/bracanalysis/}{Myriad~Genetics},
  on which it says ``People with a mutation in either the BRCA1 or
  BRCA2 gene have risks of up to 87\% for developing breast cancer by
  age 70.''  This is clearly an enumerative probability.  So was
  Angelina (or her doctor) right to interpret it as her own individual
  risk?
\end{ex}

\begin{ex}
  \label{ex:climate2}
  The events and scenarios considered by the IPCC are fundamentally
  ``one-off,'' so any attempt at interpreting the quoted probabilities
  must be individualistic.  But of what nature?  (It was suggested in
  a radio commentary that a probability of $95\%$ means that $95\%$ of
  scientists agree with the statement).
\end{ex}

\begin{ex}
  \label{ex:obama2}
  It would be very interesting to know just how the President
  conceived of this $55\%$ probability of success for a one-off event.
  We might also ask: Was that probability assessment justified (in any
  sense) by the turn-out of events?
\end{ex}

\begin{ex}
  \label{ex:civil2}
  While there has been much discussion as to relationship between
  ``legal'' and ``mathematical'' probabilities, this is apparently an
  assertion about uncertainty in an individual case.  Such assessments
  might best be construed as personal probabilities, though they are
  notoriously subject to bias and volatility \cite{fox2002}.
\end{ex}

\section{Risk and Expert Assignments}
\label{sec:expert}
Our focus in this article is on the concept of the single-case
``individual risk,'' and we shall be exploring how this is or could be
interpreted from the point of view of the various different
conceptions of probability outlined above.  One theme common to a
number of those conceptions is that of risk as an ``expert
assignment'' \cite{gaifman:1988}.  This means that, if You start from
a position of ignorance, and then somehow learn (only) that ``the
risk'' (however understood) of outcome $A$ is $p$, then $p$ should be
the measure of Your new uncertainty about $A$.

\subsection{Personal Probability}
\label{sec:persexp}

Consider this first from the Personalist point of view.  Suppose You
learn the personal probability $p$, for event $A$, of an individual
$E$ (the ``expert'') who started out with exactly the same overall
personal probability distribution as You did, but has observed more
things, so altering her uncertainty.  By learning $p$ You are
effectively learning, indirectly, all the relevant extra data that $E$
has brought to bear on her uncertainty for $A$, so You too should now
assign personal probability $p$ to event $A$.  Note that this
definition of expert is itself a personal one: an expert for You need
not be an expert for some one else with different opinions or
knowledge.

This example shows that the property of being an expert assignment is
quite weak, since there could be a number of different experts who
have observed different things and so have different updated personal
probabilities.  Whichever one of these You learn, You should now use
that value as Your own.

Regardless of which expert You are considering consulting (but have
not yet consulted), Your current expectation of her expert probability
$p$ will be Your own, unupdated, personal probability of $A$.  In
particular, mere knowledge of the existence of an expert has no effect
on the odds You should currently be willing to offer on the outcome of
$A$.

What if You learn the personal probabilities of several different
experts?  It is far from straightforward to combine these to produce
Your own revised probability \cite{ddm:1995}, since this must depend
on the extent to which the experts share common information, and would
typically differ from each individual expert assignment---even if
these were all identical.

\subsection{Chance}
\label{sec:chanceexp}

As for ``objective chance,'' the Principal Principle makes explicit
that (whatever it may be) it should act as an expert assignment, and
moreover that this should hold for every personalist---it is a
``universal expert.''  But while this constrains what we can take
objective chance to be, it is far from being a characterisation.

\subsection{Frequency Probability}
\label{sec:freqexp}

The relationship between frequencies and expert assignments is
considered in detail in \secref{g2i} below.

\section{Group to Individual Inference for Repeated Trials}
\label{sec:g2i}

As we have seen, some conceptions of Probability are fundamentally
``groupist,'' and others fundamentally ``individualist.''  That does
not mean that a groupist approach has nothing to say about individual
probabilities, nor that an individualist approach can not address
group issues.  But the journey between these two extremes, in either
direction, can be a tricky one.  Our aim in this article is to explore
this journey, with special emphasis on the interpretation of
individual probabilities, or ``risks.''  In this Section, we consider
the ``group to individual'' (G2i) journey; the opposite (i2G)
direction will be examined in \secref{i2g} below.

We start by considering a simple archetypical example.  A coin is to
be tossed repeatedly.  What is ``the probability'' that it will land
heads (H) up (event $E_1$) on the first toss ($I_1$)?  We shall
consider how various conceptions of Probability might approach this
question.

\subsection{Frequency probability}
\label{sec:freqprob2}
The frequency approach apparently has nothing to say about the first
toss $I_1$.  Suppose however that (very) lengthy experimentation with
this coin has shown that the limiting relative frequency of heads,
over infinitely many tosses, is $0.3$.  Can we treat that value $0.3$
as representing uncertainty about the specific outcome $E_1$?  Put
otherwise, can we treat the limiting relative frequency of heads as a
(universal) expert assignment for the event of heads on the first
toss?

While there is no specific warrant for this move within the theory
itself, it would generally be agreed that we are justified in doing so
if all the tosses of the coin (including toss $I_1$) can be regarded
as
\begin{quote}
  ``repeated trials of the same phenomenon under identical
  conditions.''
\end{quote}
We shall not attempt a close explication of the various terms in this
description, but note that there is a basic assumption of {\em
  identity of all relevant characteristics\/} of the different tosses.

\subsection{Personal probability}
\label{sec:persprob2}

By contrast, the Personalist You would be perfectly willing to bet on
whether or not the next toss will land heads up, even without knowing
how other tosses of the coin may turn out.  But how do Your betting
probabilities relate to frequencies?

In order to make this connexion, we have to realise that You are
supposed to be able to assess Your betting probability, not merely for
the outcome of each single toss, but for an arbitrary specified
combination of outcomes of different tosses: for example, for the
event that the results of the first 7 tosses will be HHTTHTH in that
order.  That is, You have a full personal {\em probability
  distribution\/} over the full sequence of future outcomes.  In
particular, You could assess (say) the {\em conditional\/} probability
that the 101th toss would be H, given that there were (say) 75 Hs and
25 Ts on the first 100 tosses.  Now, before being given that
information You might well have no reason to favour H over T, and so
assign {\em unconditional\/} probability close to $0.5$ to getting a H
at the 101th toss.  However, after that information becomes available
You might well favour a {\em conditional\/} probability closer to
$0.75$.  All this is by way of saying that---unlike for the Bernoulli
model for repeated trials introduced in \secref{statprob} above---in
Your joint betting distribution the tosses would typically {\em not\/}
be independent; since, if they were, no information about the first
100 tosses could change Your probability of seeing H on the 101th.

But if You cannot assume independence, what can You assume about Your
joint distribution for the tosses?  Here is one property that might
seem reasonable: that You simply do not care what order You will see
the tosses in.  This requires, for example, that Your probability of
observing the sequence HHTTHTH should be the same as that of the
sequence HTHHTHT, and the same again for any other sequence containing
4 Hs and 3 Ts.  This would not be so if, for example, You felt the
coin was wearing out with use and acquiring an increasing bias towards
H as time passes.

This property of the irrelevance of ordering is termed {\em
  exchangeability\/}.  It is much weaker than independence, and will
often be justifiable, at least to an acceptable approximation.

Now it is a remarkable fact \cite{definetti:37} that, so long only as
Your joint distribution is exchangeable, the following must hold:
\begin{enumerate}
\item
  \label{it:df1} You believe, with probability 1, that there will
  exist a {\em limiting relative frequency\/}, $p =
  \lim_{N\rightarrow\infty} f_N$, of H's, as the number $N$ of tosses
  observed tends to infinity.
\item
  \label{it:df2} You typically do not initially know the value of $p$
  (though You could place bets on that value---You have a distribution
  for $p$); but {\em if\/} You were somehow to learn the value of $p$,
  then conditionally on that information You would regard all the
  tosses as being independent, with common probability $p$ of landing
  H.  (There are echoes here of the Principal Principle).
\end{enumerate}
According to \itref{df1}, under the weak assumption of
exchangeability, the ``individualist'' Personalist can essentially
accept the ``groupist'' Frequency story: more specifically, the
Bernoulli model (with unknown probability $p$).  But the Personalist
can go even further than the Frequentist: according to \itref{df2},
exchangeability provides a warrant for equating the ``individual
risk,'' on each single toss, with the overall ``group probability''
$p$ (if only that were known\ldots).  That is, the Bernoulli model
will be agreed upon by all personalists who agree on exchangeability
\cite{apd:intersub}, and the limiting relative frequency $p$, across
the repeated trials, constitutes a universal expert assignment for
this class.

However, an important way in which this Personalist interpretation of
the Bernoulli model differs from that of the Frequentist is in the
conditions for its applicability.  Exchangeability is justifiable when
You have {\em no sufficient reason\/} to distinguish between the
various trials.  This is a much weaker requirement than the
Frequentist's ``identity of all relevant characteristics.''  For
example, suppose You are considering the examination outcomes of a
large number of students.  You know there will be differences in
ability between the students, so they can not be regarded as identical
in all relevant respects.  However, if You have no specific knowledge
about the students that would enable You to distinguish the geniuses
from the dunces, it could still be reasonable to treat these outcomes
as exchangeable.

An important caveat is that Your judgment of exchangeability is,
explicitly or implicitly, conditioned on Your current state of
information, and can be destroyed if Your information changes.  In the
above example of the students, since You are starting with an
exchangeable distribution, You believe there will be an overall
limiting relative frequency $p$ of failure across all the students,
and that if You were to learn $p$ that would be Your correct revised
probability that a particular student, Karl, will fail the exam.
Suppose, however, You were then to learn something about different
students' abilities.  This would not affect Your belief (with
probability 1) in the existence of the group limiting relative
frequency $p$---but You would no longer be able to treat that $p$ as
the individual probability of failure for Karl (who, You now know, is
particularly bright).

As an illustration of the above, suppose You are considering the
performances of the students (whom You initially consider
exchangeable) across a large number of examination papers.  Also,
while You believe that some examinations are easier than others, You
have no specific knowledge as to which those might be, so initially
regard the examinations as exchangeable.  Now consider the full
collection of all outcomes, labelled by student and examination.  As a
Personalist, You will have a joint distribution for all of these.
Moreover, because of Your exchangeability judgments, that joint
distribution would be unchanged if You were to shuffle the names of
the students, or of the examinations.  Suppose now You are interested
in ``the risk'' that Karl will fail the Statistics examination.  You
might confine attention to Karl, and use the limiting relative
frequency of his failures, across all his other examinations, as his
``risk'' of failing Statistics.  This appears reasonable since You are
regarding Karl's performances across all examinations (including
Statistics) as exchangeable.  In particular, this limiting relative
frequency of Karl's failure, across all examinations, constitutes an
expert assignment for the event of his failing Statistics.
Alternatively, You could concentrate on the Statistics examination,
and take the limiting relative frequency of failure on that
examination, by all the other students, as measuring Karl's risk of
failing.  This too seems justifiable---and supplies an expert
assignment---because You consider the performances of all the students
(including Karl) on the Statistics examination as exchangeable.
However, these two ``risks'' will typically differ---because they are
conditioned on different information.  Indeed, neither can take
account of the full information You might have, about the performances
of all students on all examinations.  Given that full information, You
should be able to assess both Karl's ability (by comparing his average
performance with those of the other students) and the difficulty of
the Statistics examination (on comparing the average performance in
Statistics with those for other examination).  Somehow or other You
need to use all this information (as well as the performances of other
students on other examinations) to come up with Your ``true'' risk
that Karl will fail Statistics---but it is far from obvious how You
should go about this.\footnote{There is in principle a way to do this
  \cite{aldous,hoover:82}, supplying a universal expert assignment for
  the class of all personalists agreeing on exchangeability both
  across students and across examinations---but it relies on advanced
  mathematics and is highly non-trivial to implement.}

We can elaborate such examples still further.  Thus suppose students
are allowed unlimited repeat attempts at each examination, and that
(for each student-examination combination) we can regard the results
on repeated attempts as exchangeable.  Then yet another interpretation
of Karl's risk of failing Statistics on some given attempt---yet
another universal expert assignment---would be the limiting relative
frequency of failure across all Karl's repeat attempts at the
Statistics examination.  This would typically differ from all the
values discussed above.

Now it can be shown that, for this last interpretation of risk, its
value would be unaffected by further taking into account all the rest
of the data on other students' performances on other examinations: it
is truly conditional on all that is (or could be) known about the
various students' performances.  Does this mean that we have finally
identified the ``true'' risk of Karl failing Statistics?

Not so fast\ldots~\@ Suppose You consider that Karl's confidence, and
hence his performance, on future attempts will be affected by his
previous results, his risk of future failure going up whenever he
fails, and down whenever he passes.  It is not difficult to make this
behaviour consistent with the exchangeability properties already
assumed \cite{hill:87}.  As a simple model, consider an urn that
initially contains 1 red ball (representing success) and 1 green ball
(representing failure).  Karl's successive performances are described
by the sequence of draws of balls from this urn, made as follows:
whichever ball is drawn is immediately replaced, together with an
additional ball of the same colour.  At each stage it is assumed the
draws are made ``at random,'' with each ball currently in the urn
being ``equally likely'' to be the next to be drawn (this being a
reasonably straightforward application of the classical
interpretation).  Thus Karl's future performance is influenced by his
past successes and failures, with each success [failure] increasing
[decreasing] the (classical) chance of success at the next attempt.

Now it can be shown that the sequence of colours drawn forms an
exchangeable process, and it follows that the relative frequency with
which a red ball is drawn, over a long sequence of such draws, will
converge to some limit $p$.  However, at the start of the process $p$
is not known, but is distributed uniformly over the unit interval.
How would you assess the probability that the first draw will result
in a red ball?  Useless to speculate about the currently unknown value
of the limiting relative frequency $p$; the sensible answer is surely
the ``classical'' value $1/2$.

In this case, while You still believe that there will exist a limiting
relative frequency of failure across all Karl's attempts at Statistics
(and this constitutes a universal expert assignment), not only is this
initially unknown to You, but its very value can be regarded as being
constructed over time, as Karl experiences successes and failures and
his level of performance gets better and worse accordingly.  So why
should You regard the limiting relative frequency of future failures,
dependent as this is on Karl's randomly varying performance in his
future attempts, as an appropriate measure of his risk of failing on
this, his first attempt?  As commented by \textcite{cane:1977} in a
parallel context: ``\ldots if several clones were grown, each under
the same conditions, an observer{\ldots}might feel that the various
values [{\em \eg, of a limiting proportion---APD\/}] they showed
needed explanation, although these values could in fact be attributed
to chance events.''  This point is relevant to the assessment of risk
in the context of an individual's criminal career (\cf \exref{arai}),
where the very act of committing a new offence might be thought to
raise the likelihood of still further offences.  ``An increase in
criminal history increases the likelihood of recidivism, and a lack of
increase can reduce that likelihood.  Because criminal history can
increase (or not) over time and each crime's predictive shelf life may
be limited, it seems important to conceptualize criminal history as a
variable marker'' \cite{monahan:14}.

It is interesting to view this ambiguity as to what should be taken as
the ``real'' risk of success---the ``classical'' proportion of red
balls currently in the urn, or the ``frequentist'' proportion of red
balls drawn over the whole sequence---from a Personalist standpoint.
Both constitute expert assignments for You, but they differ.  The
former is more concrete, in that You can actually observe (or compute)
it at any stage, which You can not do for the latter.

Suppose You are forced to bet on the colour of the first ball to be
drawn.  From the classical view, the probability value to use is $0.5$
(and that classical value is fully known to You).  Alternatively,
taking a frequency view, You would consider the unknown limiting
proportion $p$, with its uniform distribution over the unit interval.
For immediate betting purposes, the relevant aspect of this is its
expectation---which is again $0.5$.  Thus the personalist does not
have to choose between the two different ways of construing ``the
probability'' of success.  (This is a special case of the result
mentioned in \secref{persexp} above).  And this indifference extends
to each stage of the process: if there are currently $r$ red and $b$
black balls in the urn, Your betting probability for next drawing a
red, based on Your current expectation (given Your knowledge of $r$
and $b$, which are determined by the results of previous draws) of the
unknown limiting ``frequency probability'' $p$, will be
$r/(r+b)$---again agreeing with the known ``classical'' value.

But what is the relevance of the above ``balls in urn'' model to the
case of Karl's repeat attempts?  Even though the two stories may be
mathematically equivalent, there is no real analogue, for Karl, of the
``classical'' probability based on counting the balls in the urn: Your
probability $r/(r+b)$ of Karl's failing on his first attempt is merely
a feature of Your Personalist view of the world, with limited
relevance for any one else.  So---what is ``the risk'' in this
situation?

\section{Individual Risk}
\label{sec:indrisk}

Stories of coin tosses and such are untypical of real-world
applications of risk and probability.  So now we turn to a more
realistic example:

\begin{ex}
  \label{ex:sam} What is ``the risk'' that Sam will die in the next 12
  months?
\end{ex}

You might have good reason to be interested in this risk: perhaps You
are Sam's life assurance company, or You Yourself are Sam.  There is
plenty of mortality data around; but Sam is an individual, with many
characteristics that, in sum (and in Sam), are unique to him.  This
makes it problematic to apply either of the G2i arguments above.  As a
Frequentist, You would need to be able to regard Sam and all the other
individuals in the mortality data-files as ``identical in all relevant
characteristics,'' which seems a tall order; while the Personalist You
would need to be able to regard Sam and all those other individuals as
exchangeable---but will typically know too much about Sam for that
condition to be appropriate.

How then could You tune Your risk of Sam's death to the ambient data?
A common way of proceeding is to select a limited set of background
variables to measure on all individuals, Sam included.  For example,
we might classify individuals by means of their age, sex, smoking
behaviour, fruit and alcohol intake, and physical activity.  We could
then restrict attention to the subset of individuals, in the data,
that match Sam's values for these variables, and regard the (ideal
limiting) relative frequency of death within 12 months in that
subset\footnote{Or, we could set up a statistical model for the form
  of the dependence of this risk on the given attributes, estimate its
  parameters from the ambient data, and apply it to Sam.}  as a
measure of Sam's own risk.\footnote{An animation for exploring the
  dependence of this frequency on these characteristics (as well as
  calendar year) can be found at the website
  \href{http://understandinguncertainty.org/files/animations/Survival1/Survival.html}{\tt
    understandinguncertainty.org}.}  To the extent that You can regard
Sam as exchangeable with all the other individuals sharing his values
for the selected characteristics, the limiting relative frequency in
that group constitutes an expert assignment for Sam.  There can be no
dispute that. from a pragmatic standpoint, such information about
frequencies in a group of people ``like Sam'' (or ``like Angelina'')
can be extremely helpful and informative.  In \textcite{imreydawid}
such a group frequency, regarded as relevant to an individual member
of the group, is termed an ``individual{\em{ized}\/} risk.''  The
foundational philosophical question, however, is: Can we consider this
as supplying a measure of ``individual risk''?  For the Frequentist,
that would require a belief that the chosen attributes capture ``all
relevant characteristics'' of the individuals; for the Personalist, it
would require that You have no relevant additional information about
Sam (or any of the other individuals in the data), and can properly
assume exchangeability---conditional on the limited information that
is being taken into account.  Neither of these requirements is fully
realistic.

In any case, irrespective of philosophical considerations, there are
two obvious difficulties with the above ``individualization''
approach:
\begin{itemize}
\item The ``risk'' so computed will depend on the choice of background
  variables.
\item We may be obliged to ignore potentially relevant information
  that we have about Sam.
\end{itemize}
These difficulties are often branded ``the problem of the reference
class,'' and would seem to bedevil any attempt to construct an
unassailable definition of ``individual risk'' from a groupist
perspective.

\subsection{``Deep'' risk}
\label{sec:deep}
The above difficulties might disappear if it were the case that, as we
added more and more background information, the frequency value in the
matching subpopulation settled down to a limit: what we might term the
``deep'' risk, conditional on {\em all\/} there is to know about Sam.
However, it is not at all clear why such limiting stability should be
the case, nor is there much empirical evidence in favour of such a
hypothesis.  Indeed, it would be difficult to gather such evidence,
since, as we increase the level of detail in our background
information, so the set of individuals who match Sam on that
information will dwindle, eventually leaving just Sam himself.

Even for the case of tossing a coin, it could be argued that if we
know ``too much'' about the circumstances of its tossing, that would
lead to a very different assessment of the probability of H; perhaps
even, with sufficiently microscopic information about the initial
positions and momenta of the molecules of the thumb, the coin, the
table and the air, the outcome of the toss would become perfectly
predictable, and the ``deep'' probability would reduce to 0 or 1 (or
perhaps not, if we take quantum phenomena into account\ldots).

I would personally be sympathetic to the view that this {\em reductio
  ad absurdum\/} is a ``category error,'' since no one really intended
the ``deep'' information to be quite {\em that\/} deep.  Perhaps it is
the case that the addition of more and more ``appropriate'' background
information, at a more superfical level, would indeed lead to a
stabilisation of the probability of H at some non-trivial value.  But
this account is full of vagueness and ambiguities, and begs many
questions.  And even if we could resolve it for the extremely
untypical case of coin-tossing, that would not give us a licence to
assume the existence of ``deep risk'' for more typical practical
examples, such as Sam's dying in the next 12 months.

\section{A Different Approach: Individual to Group Inference }
\label{sec:i2g}

All our discussion so far---even for the ``individualist'' Personalist
conception of Probability---has centred on ``Group to Individual''
(G2i) inference: taking group frequencies as our fundamental starting
point, and asking how these might be used to determine individual
risks.  And it has to be said that we have reached no conclusive
answer to this question.  So instead we now turn things upside down,
and ask: Suppose we take individual risks as fundamental---how then
could we relate these to group frequencies, and with what
consequences?  This is the ``Individual to Group'' (i2G) approach.  As
we shall see, although we will not now be defining individual risks in
terms of group frequencies, those frequencies will nevertheless
severely constrain what we can take the individual risks to be.

Our basic framework is again an ordered population of individuals
$(I_1, I_2, \ldots)$.  For individual $I_k$, You have an outcome event
of interest, $E_k$, and some background information, $H_k$.  We denote
Your ``information base''---Your full set of background information
$(H_1, H_2, \ldots)$ on all individuals---by $\cal H$.

Note, importantly, that in this approach we need not assume that You
have similar information for different individuals, nor even that the
different outcome events are of the same type (though in applications
that will typically be the case).  In particular, we shall not impose
any analogue of either the Frequentist's condition of ``repeated
trials under identical conditions,'' or the Personalist's judgment of
exchangeability.

We start with an initial bold assumption: that You have been able to
assess, for each individual $I_k$, a probability, $p_k = \Pr(E_k \mid
H_k)$, for the associated event $E_k$, in the light of the associated
background information $H_k$.  This might, but need not be,
interpreted as a Personalist betting probability: all that we want is
that it be some sort of ``individualist'' assessment of uncertainty.
We put no other constraint on these ``individual risks''---most
important, they will typically vary from one individual to another.
We term $p_k$ a {\em probability forecast\/} for $E_k$.  For some
purposes we will require that the forecasts are based {\em only\/} on
the information in ${\cal H}$, and nothing else.  There are some
subtleties involved in making this intuitively meaningful condition
mathematically precise---one approach is through the theory of
computability \cite{apd:empprob}.  Here we will be content to note
that this can be done.  We shall term such forecasts {\em ${\cal
    H}$-based\/}.

There are two similar but slightly different scenarios that we can
analyse, with essentially identical results.  In the first, which we
may term the {\em independence\/} scenario, the background information
$H_k$ pertains solely to individual $I_k$, and it is supposed that,
given $H_k$, Your uncertainty about $E_k$, as expressed by $p_k$,
would not change if You were to receive any further information (be it
background information or outcome information) on any of the other
individuals.  In the second, {\em sequential\/} scenario, $H_k$
represents the total background information on all previous and
current individuals $I_1, \ldots, I_k$, as well as the outcomes of
$E_1, \ldots, E_{k-1}$ for the previous individuals; no other
conditions need be imposed.

An important practical application, in the sequential formulation, is
to weather forecasting, where $I_k$ denotes day $k$, $E_k$ denotes
``rain on day $k$,'' $H_k$ denotes the (possibly very detailed)
information You have about the weather (including whether or not it
rained) up to and including the previous day $k-1$, and You have to go
on TV at 6pm each evening and announce Your probability $p_k$ that it
will rain the following day $I_k$.  We will often use this particular
example to clarify general concepts.

Our proposed relationship between individual probabilities and group
frequencies will be based on the following idea: Although You are free
to announce any probability values You want for rain tomorrow, if You
are to be trusted as a reliable weather forecaster, these values
should bear some relationship to whether or not it does actually rain
on the days for which You have issued forecasts.  Note that this
approach judges Your probabilities by comparison with the outcomes of
the events---not with the ``true probabilities'' of the events.  No
commitment as to the existence of ``true probabilities'' is called
for.

\subsection{Calibration}
\label{sec:simpcal}
Let $e_k$ denote the actual outcome of event $E_k$, coded $1$ if $E_k$
happens, and $0$ if it does not.

\subsubsection{Overall calibration}
\label{sec:overallcal}
We start by proposing the following {\em overall calibration
  criterion\/} of agreement between the probability forecasts $(p_k)$
and the outcomes $(e_k)$: over a long initial sequence $I_1, I_2,
\dots, I_N$, the overall proportion of the associated events that
occur,
\begin{equation}
  \label{eq:prop1}
  \frac{\sum_{i=1}^Ne_{i}}N,
\end{equation}
which is a ``groupist'' property, should be close to the average of
the ``individualist'' forecast probabilities,
\begin{equation}
  \label{eq:av1}
  \frac{\sum_{i=1}^Np_{i}}N.
\end{equation}

This seems a plausible requirement, but do we have a good warrant for
imposing it?  Yes.  It can be shown \cite{apd:wellcal} that (for
either of the scenarios) the overall calibration property is assigned
probability 1 by Your underlying probability distribution.  That is to
say, You firmly believe that Your probability forecasts will display
overall calibration.  So, if overall calibration turns out {\em not\/}
to be satisfied, an event that You were convinced was going to happen
has failed to occur---a serious anomaly, that discredits Your whole
distribution, and with it Your probability forecasts
\cite{apd:synthesis}.  Overall calibration thus acts as a minimal
``sanity check'' on your probabiity forecasts: if it fails, You are
clearly doing something wrong.

However, that does not mean that, if it holds, You are doing
everything right: overall calibration is a weak requirement.  For
example, in an environment where it rains $50\%$ of the time, a
weather forecaster who, ignoring all information about the past
weather, always announces a probability of $50\%$, will satisfy
overall calibration---but if in fact it rains every alternate day (and
much more generally) he will be showing little genuine ability to
forecast the changing weather on a day-by-day basis.  Another
forecaster, who has a crystal ball, always gives probability 1 or 0 to
rain, and always gets it right.  Her perfect forecasts also satisfy
overall calibration.

In order to make finer distinctions between forecasts of such very
different quality, we will progressively strengthen the calibration
criterion, through a number of stages.  Note, importantly, that every
such strengthened variant will share the ``sanity check'' function
described above for overall calibration: according to Your probability
distribution, it will be satisfied with probability 1.  So, if it
fails, Your probability forecasts are discredited.

\subsubsection{Probability calibration}
\label{sec:probcal}
For our next step, instead of taking the averages in \eqref{prop1} and
\eqref{av1} over all days until day $N$, we focus on just those days
$I_k$ for which the forecast probability $p_k$ was equal to (or very
close to) some pre-assigned value.  If that value is, say, $30\%$ (and
that value is eventually used infinitely often) then {\em probability
  calibration\/} requires that, in the limit, the proportion of these
days on which it in fact rains should be (close to) $30\%$; and
similarly for any other pre-assigned value.

However, although probability calibration is again a very natural
idea, it is still too weak for our purpose.  In particular, it will
still be satisfied for both our above examples of uninformative
forecasts and of perfect forecasts.

\subsubsection{Subset calibration}
\label{sec:subsetcal}

For our next attempt, we allow the averages in \eqref{prop1} and
\eqref{av1} to be restricted to a subset of the individuals,
arbitrarily chosen except for the requirement that it must be selected
without taking any account of the values of the $E$ and the $H$'s.
For example, we might choose every second day.  If in fact the weather
alternates wet, dry, wet, dry,\ldots, then the uninformative
forecaster, who always says $50\%$, will now fail on this criterion,
since if we restrict to the odd days alone his average forecast
probability is still $50\%$, but the proportion of rainy days will now
be $100\%$.  The perfect forecaster will however be announcing a
probability forecast of $100\%$ for every odd day, and so will satisfy
this criterion.

However, although subset calibration has succeeded in making the
desired distinction, even this is not strong enough for our purposes.

\subsubsection{Information-based calibration}
\label{sec:infcal}
In all our attempts so far, we have not made any essential use of the
background information base ${\cal H}$, and the requirement that the
forecasts be ${\cal H}$-based.  But we cannot properly check whether a
forecaster is making appropriate use of this background information
without ourselves taking account of it.

In the sequential weather forecasting scenario, the forecaster is
supposed to be taking account of (at least) whether or not it rained
on previous days, and to be responding appropriately to any pattern
that may be present in those outcomes.  To test this, we could form a
test subset in a dynamic way, ourselves taking account of all the
forecaster's background information (but, to be totally fair to the
forecaster, nothing else\footnote{Again, this requirement can be
  formalised using computability theory.}).  Thus we might consider,
for example, the subset comprising just those Tuesdays when it had
rained on both previous days.  If the forecaster is doing a proper
job, he should be calibrated (\ie, his average probability forecast
should agree with the actual proportion of rainy days), even if we
restrict the averages to be over such an ``${\cal H}$-based'' subset.
An essentially identical definition applies in the independence
scenario.

When this property is satisfied for all ${\cal H}$-based subsets, we
will call the probability forecasts {\em ${\cal H}$ calibrated\/}.  A
set of forecasts that is both ${\cal H}$-based and ${\cal
  H}$-calibrated will be called {\em ${\cal H}$-valid\/}.  Again we
stress that You assign probability 1 to Your $\cal H$-based forecasts
being $\cal H$-valid, so this is an appropriate condition to impose on
them.

Finally, we have a strong criterion relating individual probability
forecasts and frequencies.  Indeed, it constrains the individuals
forecasts so much that, in the limit at least, their values are fully
determined.  Thus suppose that we have two forecasters, who issue
respective forecasts $(p_k)$ and $(q_k)$, and that both sets are
${\cal H}$-valid.  It can then be shown \cite{apd:wellcal} that, as
$k$ increases without limit, the difference between $p_k$ and $q_k$
must approach $0$.  We may term this result {\em asymptotic
  identification\/}.

This is a remarkable result.  We have supposed that ``individual
risks'' are given, but have constrained these only through the
``groupist'' ${\cal H}$-calibration criterion.  But we see that this
results ``almost'' in full identification of the individual values, in
the sense that, if two different sets of probability forecasts both
satisfy this criterion, then they must be essentially identical.  In
particular, if there exists any set of $\cal H$-valid
forecasts,\footnote{which is however not guaranteed
  \cite{schervish:85,belot}} then those values are the ``essentially
correct'' ones: any other set of $\cal H$-based forecasts that does
not agree, asymptotically, with those values can not be $\cal
H$-valid---and is thus discredited.  So in this sense the i2G approach
has succeeded---where the more traditional G2i approaches failed---in
determining the values of individual risks on the basis of group
frequencies.

As an almost too simple example of this result, in the independence
scenario, suppose that $H_k$, the background information for
individual $I_k$, is his score on the Violence Risk Appraisal Guide
ARAI \cite{VRAG}, with 9 categories.  If this limited information
${\cal H}$ is all that is available to the forecaster, his forecasts
will be ${\cal H}$-based just when they announce the identical
probability value for all individuals in the same VRAG category.  And
they will be ${\cal H}$-calibrated if and only if, within each of the
9 categories, his announced probability agrees with the actual
reoffending rates.  That is, ${\cal H}$-validity here reduces to the
identification of {\em individual\/} risk with {\em individualized\/}
risk; and this identification is thus justified so long as Your
complete information is indeed restricted to that comprised by ${\cal
  H}$.

What this simple example fails to exhibit is that, in more general
cases, the ``determination'' of the individual forecasts is only
asymptotic: given a set of $\cal H$-valid forecasts, we could
typically change their values for a finite number of individuals,
without affecting ${\cal H}$-validity.  So we are not, after all, able
to associate a definitive risk value with any single individual.

Another big downside of the above result is that, while assuring us of
the essential uniqueness of $\cal H$-valid forecasts, it is entirely
non-constructive.  It seems reasonable to suppose that, once we know
that such uniquely determined forecasts exist, there should be a way
to construct them: for example, it would be nice to have an algorithm
that would issue probability forecasts for the following day, based on
past weather (${\cal H})$, that will be properly calibrated, however
the future weather in fact turns out.  But alas!\ in general this is
not possible, and there is no ${\cal H}$-based system that can be
guaranteed to be ${\cal H}$-calibrated \cite{oakes:1985,apd:oakes}.

Another potentially serious limitation of the i2G approach is its
dependence on the population of individuals considered, and moreover
on the order in which they are strung out as a single sequence.  It
does not seem easy to accommodate a structure, such as considered in
\secref{persprob2}, where many students take many resits of many
examinations.  For one thing, it seems problematic to incorporate the
judgements of exchangeability made there into the i2G approach; for
another, different ways of forming a sequence of
student-examination-resit combinations could well lead to mutually
inconsistent calibration requirements.

\subsection{Varying the information base}
\label{sec:vary}
A fundamental aspect of the i2G approach to individual risk is that
this is a relative, not an absolute, concept: its definition and
interpretation depend explicitly on the information that is being
taken into account,\footnote{In this it has some of the flavour of
  Logical Probability} as embodied in ${\cal H}$.  If we change the
information base, the associated valid probability values will change.

We can relate the i2G risks based on different information bases, one
more complete than the other.  Thus suppose that ${\cal K} = (K_k)$ is
an information base that is more detailed than ${\cal H}$: perhaps
$H_k$ only gives information about whether or not it rained on days
prior to day $k$, while $K_k$ also contains information about past
maximum temperature, wind speed, \etc\@ Suppose $(p_k)$ is a set of
${\cal H}$-valid forecasts, and $(q_k)$ a set of ${\cal K}$-valid
forecasts.  These would typically differ, even asymptotically: We
would expect the $(q_k)$, being based on more information, to be
``better'' than the $(p_k)$, and so would {\em not\/} expect $p_k -
q_k \rightarrow 0$.  This does not contradict asymptotic
identification.  The $(q_k)$ are not ${\cal H}$-based, so not ${\cal
  H}$-valid.  Also, while the $(p_k)$ are ${\cal K}$-based, they have
not been required to be ${\cal K}$-calibrated, so they are not ${\cal
  K}$-valid.  So, whether we take the underlying information base to
be ${\cal H}$ or ${\cal K}$, the conditions implying asymptotic
identification simply do not apply.

It does however follow from the argument for asymptotic identification
that, if we consider the subsequence for which, say, $p_k = 0.4$ (to a
good enough approximation), then---while we can expect the $(q_k)$ in
that subsequence to vary (because of variations in the additional
information in ${\cal K}$ that they take into account)---their
limiting average value in the subsequence will be $0.4$ (and similarly
for any other target value for the $(p_k)$).  In this sense the
``deeper'' risks $(q_k)$ can be regarded as varying ``randomly'' about
the ``shallower'' risks $(p_k)$---just as the actual outcomes $(E_k)$
do.  Some of the attempts to model individual risks as random, such
that of \textcite{jr+sg:89}, might be understood as contemplating an
expanded but unobservable information base ${\cal K}$, and
interpreting the ``true risks'' as the ${\cal K}$-valid ones.
However, with access only to the shallower information base ${\cal
  H}$, we can only ``observe'' certain crude averages of these ``true
risks.''  If ${\cal H}$ is indeed all the information at our disposal,
there is nothing of value to be gained by extending consideration to
the unobservable deeper risks $(q_k)$---which in any case depend on
the essentially arbitrary specification of the unobserved deeper
information base ${\cal K}$.  The flip-side of this is that, when we
do have access to the deeper information, we should use that to
identify the relevant ``individual risk,'' rather than be satisfied
with a cruder average value based on more superficial information.

% Sometimes there may be a reasonably natural specification for the
% deep information base ${\cal K}$.  Thus in a case, as considered in
% \secref{persprob2}, of exchangeable students taking a Statistics
% examination just once, ${\cal H}$ is essentially empty, leading to a
% ``shallow'' assessment of the risk that student~$i$ will fail that
% agrees with the group risk, $p$ say: the overall proportion of
% students who fail.  But we might now include, in the deeper
% information base ${\cal K}$, hypothetical information about
% students' performances on further resits of the examination---which
% would equate student~$i$'s ``deep'' risk of failure, $q_i$, with his
% proportion of failures in resits.  This is essentially the
% suggestion made by \textcite[\S~3/2]{imreydawid} as to how
% ``individual risk'' might be interpreted in the context of the
% debate about ARAIs.  However we have already argued that this
% proportion need not be an appropriate measure of a student's risk on
% his first attempt.  In any case, in the absence of any data relating
% to repeat attempts, there is nothing that could be learned about
% these deep risks $(q_i)$, beyond the fact that their overall
% average, across all students, must agree with the (observable) group
% risk $p$.  That is to say, all we can learn about are the ${\cal
% H}$-valid forecasts---so why even bother to elaborate the story
% beyond these?  See \S~3.2 \todo{``Intervals for individual risks''
% ---check final numbering} of \textcite{imreydawid} for further
% discussion of this point from a different but related perspective.

An interesting implication of the above argument is that our
``individual-to-group'' approach allows for the assignment of
non-extreme probabilities to events, even when we believe in ``deep
determinism''---that is, we believe that, given a suitably detailed
information base ${\cal K}$, it would be possible to forecast the
future perfectly (in which case ${\cal K}$-valid probabilities $(q_k$)
would have to be, asymptotically, $0$ or $1$).  However, if we only
have access to a less detailed information base ${\cal H}$, the
associated ${\cal H}$-valid risk values would normally be non-extreme.
This approach thus justifies the use of probability as a description
of a system we believe to operate deterministically, so long as we are
in ignorance of the deep determining circumstances.

\section{Discussion}
\label{sec:disc}

I have surveyed a number of conceptions of the meaning of ``individual
risk,'' and found all of them wanting to some degree.  The various G2i
approaches all founder on the ``problem of the reference class,''
which does not have an unambiguous solution---although reasonable
pragmatic choices can often be made, and defended as such.  The i2G
approach avoids ambiguity and delivers individual risk values, nicely
calibrated to the information considered available---but only ``at
infinity,'' and even then these asymptotic values are uncomputable.
It also does not seem able to take account of sophisticated
exchangeability requirements.

What then is one to do?  Should the whole idea of individual risk be
abandoned?  I think this is too extreme, but certainly the concept and
nomenclature should not be bandied about carelessly, and it behoves
any one using the term to give an account of what they (think they)
mean by it.

For what it is worth, my own tentative attitude is as follows, based
on an essentially personalistic viewpoint.  Suppose I am tasked with
assessing ``the risk'' that Cain, a prisoner up for possible parole,
will if released commit a violent act (say, within the next 24
months).  Then I should think about Cain, and, taking full account of
all I know about him and others like him (and even unlike him),
compute my probability forecast by assessing the odds at which I would
be willing to bet that he will commit a violent act.  This probability
can in principle be considered as formed, by conditioning on the
information I have on Cain and others, from my full personal joint
probability distribution for the properties and outcomes of all these
individuals.  Under certain strong conditions on that joint
distribution, such as exchangeability, my forecast probability would
be close to the proportion of such events in people ``like Cain''; but
more generally, that proportion, and other relevant proportions, would
inform, but not directly constrain, my personal, properly conditioned,
individual probability forecast for Cain.

% Sometimes it might be interesting to speculate as to what my revised
% odds would be, if I were to learn further, ``deeper,'' information
% about Cain.  That can be done in principle by adding that
% hypothetical information to my conditioning dataset.  Since the
% deeper information is currently unknown to me, and therefore
% regarded as random, so too will be the ``deep(er) risk'' value that
% emerges from this exercise.  It might be helpful to others for me to
% announce my distribution for this deeper risk.  But it is essential
% that both they and I appreciate the dependence of this risk on the
% specific nature of the additional deep information considered, and
% are very clear as to what that is.

But, you object, what of objectivity?  You want to know ``the risk''
that Cain will be violent.  Why should you care about my personal
probability assessment?

Well, in the large my probability forecasts can be tested against
realised outcomes \cite{apd:enc(probfore)}, \eg, using finite-data
analogues of calibration \cite{fs/apd} to see how closely they align
with actual frequencies in appropriately selected sets.\footnote{One
  criterion I should definitely {\em not\/} be judged on is how good
  my probability forecasts are as estimates of the ``objective''
  individual risks.  This is for two reasons: first, in the
  unavoidable absence of knowledge of the values of those risks, this
  can not be done; and, secondly, there are no such things!}  If, by
such tests (formal or informal), my announced probability forecasts
are shown to be out of line with reality, you have every right to
discount my risk assessment for Cain.  But if they have passed a
suitable battery of such tests---are, provisionally, ``valid''---then
(arguing informally by analogy with our asymptotic identification
result) we might expect them to be close to the valid risk assessments
of others.  So, in the presence of a sufficient quantity of relevant
data to allow us to conduct such statistical tests, you should be able
to judge whether my risk assessments are reasonably ``objective,'' and
if so have some limited confidence in my---now statistically
justified---announced risk for Cain.

But recall that ``valid'' probability forecasts depend on the
information that is being taken into account.  Ideally this should be
the most detailed information we have about Cain and those like him;
but, the more detailed the information is, the harder it will be for
me to make my forecasts valid.\footnote{Individual risk assessment
  based on very detailed personal information becomes essentially a
  matter of ``clinical judgment.''  Although practitioners in many
  disciplines---medicine, law, psychiatry, \etc~\etc---often have
  great confidence in their own clinical judgments, these can be very
  far from being valid \cite{meehl}.}  So I might choose to
artifically restrict my information base, perhaps to just a few simple
characteristics.  I would thus be sacrificing incisiveness---the
possibility (however remote) of making ``deep'' probability forecasts,
valid with respect to the deep information that I could, in principle,
take into account---in favour of robustness---the enhanced prospect of
achieving validity with respect to deliberately restricted
information.  In the judicial context, and many others, this sacrifice
might well be considered worthwhile: indeed, a suitable specification
of what would be ``appropriate'' characteristics to take into account
could be enshrined in statute.\footnote{This leaves the possibility
  that an individual might appeal on the grounds that relevant
  information, specific to his case, was left out of consideration.
  But it would be for the appellant to make a good case for this.}
That done, suitable statistical methodology applied to these could
produce valid risk assessments---albeit relating to a shallower level
of information than would ideally be desirable.  The various ARAI
systems that have been developed can be thought of as addressing this
task.

\section*{Acknowledgments}
I am grateful to David Faigman, Alan H\'ajek, Peter Imrey, John
Monahan, Russell Poldrack and David Spiegelhalter for helpful
comments, and to the MacArthur Foundation for support through its
Research Network on Law and Neuroscience.

\bibliographystyle{oupvar}
\bibliography{strings,StatViol,indrisk,causal,dna,allclean,afms,apdpubs}

\end{document}

%% file: bknot.tex
\newcommand{\ie}{{\em i.e.\/}\xspace}
\newcommand{\eg}{{\em e.g.\/}\xspace}

\newcommand{\cf}{{\em cf.\/}\xspace}
\newcommand{\etc}{{\em etc.\/}\xspace}

\newcommand{\HIDE}[1]{ }
\newcommand{\COMMENT}[1]{ }
\newcommand{\eqref}[1]{\mbox{(\ref{eq:#1})}}

\newcommand{\secref}[1]{\mbox{\S$\,$\ref{sec:#1}}}

\newcommand{\itref}[1]{\mbox{\ref{it:#1}}}

\newcommand{\exref}[1]{\mbox{Example~\ref{ex:#1}}}

\newtheorem{expl}{Example}%[section]
\newtheorem{definer}{Definition}%[section]
\newtheorem{algor}{Algorithm}%[section]

\newtheorem{rem*}{Remark}%[section]
\newcommand{\halm}{\hspace*{\fill} $\Box$\par}

\newenvironment{ex}{\begin{expl}\rm}{\halm\end{expl}}